\documentclass[graybox]{svmult}
\def\ps@headings{%
\def\@oddhead{\mbox{}\scriptsize\rightmark \hfil \thepage}%
\def\@evenhead{\scriptsize\thepage \hfil \leftmark\mbox{}}%
\def\@oddfoot{}%
\def\@evenfoot{}}
\usepackage{epsfig, epsf, array, latexsym, graphics, multirow, color}
\usepackage{graphicx}
\usepackage{algorithm2e}
\usepackage{amsmath}

\usepackage{helvet}         
\usepackage{courier}        
\usepackage{type1cm}        
\usepackage{makeidx}         
\usepackage{graphicx}        
\usepackage{multicol}        
\usepackage[bottom]{footmisc}

\newcommand {\mymarginpar}[1]{\marginpar{#1}}
\renewcommand {\marginpar}[1]{} 

\def\_{\rule{.3em}{.15ex}}      

\newcommand{\ls}[1]
   {\dimen0=\fontdimen6\the\font
    \lineskip=#1\dimen0
    \advance\lineskip.5\fontdimen5\the\font
    \advance\lineskip-\dimen0
    \lineskiplimit=.9\lineskip
    \baselineskip=\lineskip
    \advance\baselineskip\dimen0
    \normallineskip\lineskip
    \normallineskiplimit\lineskiplimit
    \normalbaselineskip\baselineskip
    \ignorespaces
   }

\newcommand {\bearn}{\begin{eqnarray*}}
\newcommand {\eearn}{\end{eqnarray*}}
\newcommand {\barr}{\begin{array}}
\newcommand {\earr}{\end{array}}

\newcommand {\G}{{\cal G}}

\newcommand {\N}{{\cal N}}

\def\defeq{\stackrel{\scriptstyle\rm def}{=}}

\newtheorem{assumption}[definition]{Assumption}

\newcommand {\benum} {\begin{enumerate}}
\newcommand {\eenum} {\end{enumerate}}

\newcommand {\bdesc} {\begin{description}}
\newcommand {\edesc} {\end{description}}

\newcommand {\bfig}[2] {\begin{figure}[htbp]
                        \centerline {
                         \epsfig{figure={#1},clip=,width={#2}}}}
\newcommand {\brotatefig}[2] {\begin{figure}[htbp]
                        \centerline {
                         \epsfig{figure={#1},clip=,angle=-90,width={#2}}}}

\newcommand {\bfigfirst}[2] {\begin{figure}[h]
                        \centerline {
                        \setlength{\epsfxsize}{#2}
                        \epsffile{#1}}}
\newcommand {\efig}[2]{ \caption{#2}
                        \label{fig:#1}
                        \end{figure}
                        \mymarginpar{fig:#1}}
\newcommand {\erotatefig}[2]{ \caption{#2}
                        \label{fig:#1}
                        \end{figure}
                        \mymarginpar{fig:#1}}
\newcommand {\rfig}[1]{Figure \ref{fig:#1}}

\newcommand {\btab}[1]{
                       \begin{table}
                       \centering
                       \begin{tabular}{#1}}
\newcommand {\etab}[3] {
                       \end{tabular}
                       \caption[#3]{#2}
                       \label{tab:#1}
                       \end{table}
                       \mymarginpar{tab:#1}
                       \vspace{.1in}}

\newcommand {\btabular}[1]{\begin{center}
                       \begin{tabular}{#1}}
\newcommand {\etabular}{\end{tabular}
                       \end{center}}

\newcommand {\bdefin}[1]{\begin{definition}
                      \mymarginpar{def:#1}
                      \label{def:#1} }
\newcommand {\edefin}       {\end{definition}}

\newcommand {\bassum}[1]{\begin{assumption}
                      \mymarginpar{ass:#1}
                      \label{ass:#1} }
\newcommand {\eassum}       {\end{assumption}}

\newcommand {\bpro}[1]{\begin{property}
                      \mymarginpar{pro:#1}
                      \label{pro:#1} }
\newcommand {\epro}   {\end{property}}

\newcommand {\bprop}[1]{\begin{proposition}
                      \mymarginpar{prop:#1}
                      \label{prop:#1} }
\newcommand {\eprop}       {\end{proposition}}
\newcommand {\rprop}[1]{Proposition \ref{prop:#1}}

\newcommand {\blem}[1]{\begin{lemma}
                      \mymarginpar{lem:#1}
                      \label{lem:#1} }
\newcommand {\elem}   {\end{lemma}}
\newcommand {\rlem}[1]{Lemma \ref{lem:#1}}

\newcommand {\bthe}[1]{\begin{theorem}
                      \mymarginpar{the:#1}
                      \label{the:#1} }
\newcommand {\ethe}   {\end{theorem}}
\newcommand {\rthe}[1]{Theorem \ref{the:#1}}

\newcommand {\bproof}{\noindent {\bf Proof.} \ }
\newcommand {\eproof} {\hfill \squares \\ \vspace{.3cm}}
\newcommand {\bcor}[1]{\begin{corollary}
                      \mymarginpar{cor:#1}
                      \label{cor:#1} }
\newcommand {\ecor}   {\end{corollary}}
\newcommand {\rcor}[1]{Corollary \ref{cor:#1}}

\newcommand {\bax}[1]{\begin{axiom}
                      \mymarginpar{ax:#1}
                      \label{ax:#1} }
\newcommand {\eax}       {\vspace{-.1in} \end{axiom}}

\newcommand {\bconj}[1]{\begin{conjecture}
                      \mymarginpar{conj:#1}
                      \label{conj:#1} }
\newcommand {\econj}       {\vspace{-.1in} \end{conjecture}}

\newcommand {\bex}[2]{\vspace{.1in}
                      \begin{example}
                      \mymarginpar{ex:#1}
                       {\bf #2}
                      \label{ex:#1} \em}
\newcommand {\eex}       {\end{example} \vspace{.3cm} }

\newcommand {\brem}[1]{\begin{remark}
                      \mymarginpar{rem:#1}
                      \label{rem:#1} \em }
\newcommand {\erem}   {\end{remark}}

\newcommand {\beq}[1]{\mymarginpar{eq:#1}
                      \begin{equation}
                      \label{eq:#1} }

\newcommand {\beqno}[1]{\mymarginpar{eq:#1}
                      \begin{eqnarray}
                      \nonumber}

\newcommand {\eeq}       {\end{equation}}
\newcommand {\eeqno}       { && \end{eqnarray}}
\newcommand {\req}[1]{(\ref{eq:#1})}

\newcommand {\bear}[1]{\mymarginpar{eq:#1}
                       \begin{eqnarray}
                       \label{eq:#1} }

\newcommand {\bearno}[1]{\mymarginpar{eq:#1}
                       \begin{eqnarray}
                       \nonumber}

\newcommand {\eear}{\end{eqnarray}}
\newcommand {\eearno}{\end{eqnarray}}
\newcommand {\bsel}{\left \{ \begin{array}{cl}}
\newcommand {\esel}{\end{array} \right.}

\newcommand {\bmat}[1]{\left [ \begin{array}{#1}}
\newcommand {\emat}{\end{array} \right ]}
\newcommand {\bsec}[2]{\mymarginpar{sec:#2}
                       \section{#1}
                       \label{sec:#2} }

\newcommand {\rsec}[1]{Section \ref{sec:#1}}

\def\R{I\kern-0.30em R}
\def\N{I\kern-0.30em N}
\def\P{I\kern-0.30em P}

\newcommand\squares{\vrule height6pt width7pt depth1pt}

\newcommand {\bxfig}[2] {\begin{figure}[htbp]
                        \centerline {
                         \includegraphics[width=#2]{#1}}}
\newcommand {\brotatexfig}[2] {\begin{figure}[htbp]
                        \centerline {
                         \includegraphics[width=#2,angle=90]{#1}}}
\DeclareGraphicsExtensions{.pdf,.jpg,.png}

\def\ex{{\bf\sf E}}

\def\bfa{{\mbox{\boldmath $a$}}}
\def\bfsa{{\mbox{\boldmath\scriptsize $a$}}}
\def\bfalpha{{\mbox{\boldmath $\alpha$}}}
\def\bfsalpha{{\mbox{\boldmath\scriptsize $\alpha$}}}

\def\bfeta{{\mbox{\boldmath $\eta$}}}
\def\bftheta{{\mbox{\boldmath $\theta$}}}

\def\bfc{{\mbox{\boldmath $c$}}}

\def\bff{{\mbox{\boldmath $f$}}}

\def\bfu{{\mbox{\boldmath $u$}}}
\def\bfv{{\mbox{\boldmath $v$}}}
\def\bfw{{\mbox{\boldmath $w$}}}
\def\bfx{{\mbox{\boldmath $x$}}}
\def\bfy{{\mbox{\boldmath $y$}}}
\def\bfz{{\mbox{\boldmath $z$}}}
\def\bfsx{{\mbox{\boldmath\scriptsize $x$}}}

\def\bfA{{\mbox{\boldmath $A$}}}

\def\bfD{{\mbox{\boldmath $D$}}}

\def\bfF{{\mbox{\boldmath $F$}}}
\def\bfG{{\mbox{\boldmath $G$}}}
\def\bfH{{\mbox{\boldmath $H$}}}
\def\bfI{{\mbox{\boldmath $I$}}}
\def\bfJ{{\mbox{\boldmath $J$}}}
\def\bfM{{\mbox{\boldmath $M$}}}

\def\bfone{{\mbox{\boldmath $1$}}}

\def\bfzero{{\mbox{\boldmath $0$}}}

\newcommand{\er}{Erd\H{o}s and R\'{e}nyi\ }

\begin{document}
\title*{A Generalized Configuration Model with Degree Correlations and Its Percolation Analysis}
\author{Duan-Shin Lee \inst{1}\and
Cheng-Shang Chang\inst{2}\and Miao Zhu \inst{3} \and Hung-Chih Li\inst{4}}

\institute{Institute of Communications Engineering,National Tsing Hua University,Hsinchu 300, Taiwan, R.O.C. \texttt{lds@cs.nthu.edu.tw}
\and Institute of Communications Engineering,National Tsing Hua University,Hsinchu 300, Taiwan, R.O.C. \texttt{cschang@ee.nthu.edu.tw}
\and Institute of Communications Engineering,National Tsing Hua University,Hsinchu 300, Taiwan, R.O.C. \texttt{xmzhumiao11111@hotmail.com}
\and Institute of Communications Engineering,National Tsing Hua University,Hsinchu 300, Taiwan, R.O.C. \texttt{g9964548@oz.nthu.edu.tw}}


\maketitle

\textbf{Abstract} In this paper we present a generalization of the classical configuration
model.  Like the classical configuration model, the generalized
configuration model allows users to specify an arbitrary degree
distribution.  In our generalized
configuration model, we partition the stubs in the configuration model
into $b$ blocks of equal sizes and choose a permutation
function $h$ for these blocks.
In each block, we randomly designate a number proportional
to $q$ of
stubs as type 1 stubs, where $q$ is a parameter in
the range $[0, 1]$.  Other stubs are designated as type 2 stubs.
To construct a network, randomly select an
unconnected stub.  Suppose that this stub
is in block $i$. If it is a type 1 stub, connect
this stub to a randomly selected unconnected type 1 stub in block $h(i)$.
If it is a type 2 stub, connect it to a randomly selected
unconnected type 2 stub. We repeat this process until
all stubs are connected.  Under an assumption, we derive a closed form
for the joint degree distribution of two random neighboring
vertices in the constructed graph.
Based on this joint degree distribution, we show that the Pearson
degree correlation function is linear in $q$ for any fixed $b$.
By properly choosing $h$, we show that our construction
algorithm can create assortative networks as well as
disassortative networks. We present a percolation analysis of
this model.
We verify our results by extensive computer simulations.

{\bf keywords}: configuration model, assortative mixing,
degree correlation

\section{Introduction}

Recent advances in the study of networks that arise in field of
computer communications, social interactions, biology, economics,
information systems, {\em etc.}, indicate that these seemingly
widely different networks possess a few common properties.
Perhaps the most extensively studied properties are power-law
degree distributions [1], the small-world property
[29], network transitivity or "clustering" [29].
Other important research subjects on networks include
network resilience, existence of community structures, synchronization,
spreading of information or epidemics.  A fundamental
issue relevant to all the above research
issues is the correlation between properties of neighboring vertices.
In the ecology and epidemiology
literature, this correlation between neighboring vertices is called
assortative mixing.

In general, assortative mixing is a concept that attempts to describe
the correlation between properties of two connected vertices.
Take social networks for example.  vertices may have ages, weight, or
wealthiness as their properties.  It is found that friendship between
individuals are strongly affected by age, race, income, or
languages spoken by the
individuals.  If vertices with similar properties
are more likely to be connected
together, we say that the network shows assortative mixing.  On the other
hand, if vertices with different properties are likely to be connected
together, we say that the network shows disassortative mixing.
It is found that social networks tend to show assortative mixing, while
technology networks, information networks and biological networks tend
to show disassortative mixing [21].
The assortativity level of a network is commonly measured by a quantity
proposed by Newman [20] called assortativity coefficient.
If the assortativity level to be measured is degree,
assortativity coefficient reduces to
the standard Pearson correlation coefficient [20].  Specifically,
let $X$ and $Y$ be the degrees of a pair of randomly selected
neighboring vertices, the
Pearson degree correlation function is
the correlation coefficient of $X$ and $Y$, {\em i.e.}
\beq{def-rho}
\rho(X,Y)\defeq \frac{\ex(XY)-\ex(X)\ex(Y)}{\sigma_X \sigma_Y},
\eeq
where $\sigma_X$ and $\sigma_Y$ denote the standard deviation
of $X$ and $Y$ respectively.
We refer the reader to
[20,21,12,30,22,3,17] for more information on
assortativity coefficient and other related measures.
In this paper we shall
focus on degree as the vertex property.

Researchers have found that assortative mixing plays a crucial role
in the dynamic processes, such as information or disease spreading,
taking place on the topology defined by the network
[18,3,7,17,26].
Assortativity also has a fundamental impact to the network resilience
as the network loses vertices or edges [28].
In order to study information propagation or network resilience,
researchers may need to build models with assortative mixing or
disassortative mixing.  Newman [20] and
Xulvi-Brunet {\em et al.} [30] proposed algorithms to generate
networks with assortative
mixing or disassortative mixing based on an idea of rewiring edges.
Bogu\~{n}\'{a} {\em et al.} [3] proposed a class
of random networks in which hidden variables are associated with the
vertices of the networks.  Establishment of edges are controlled by the
hidden variables.  Ramezanpour {\em et al.} [23]
proposed a graph
transformation method to convert a configuration model into a
graph with degree correlations and non-vanishing clustering coefficients
as the network grows in size.  However, the degree distribution no longer
remains the same as the network is transformed.
Zhou [31] also proposed a method to generate
networks with assortative mixing or disassortative mixing using a
Monte Carlo sampling method.  Comparing with these methods,
our method has an advantage that specified degree distributions
are preserved in the constructed networks.  In addition, our method
allows us to derive a closed form for the Pearson degree
correlation function for two random neighboring vertices.

In this paper we propose a method to generate random networks
that possess either assortative mixing property or disassortative
mixing property.  Our method is based on a modified construction
method of the configuration model proposed by Bender {\em et al.}
[2] and Molloy {\em et al.} [15].
The modified construction method is as follows.  Given a degree
distribution, generate a sequence of degrees.  Each vertex is
associated with a set of ``stubs" where the number of stubs
is equal to the degree of the vertex.  We sort and arrange the
stubs of all vertices in ascending order (descending
order would be fine as well).  Sort and
divide the stubs into $b$ blocks.  We associate each block
with another block.  This association forms a permutation, {\em
i.e.} no two distinct blocks are associated with a common block.
We randomly designate a fixed number of stubs in each block
as type-1 stubs.  The rest stubs are all designated as type-2
stubs.  To connect stubs, we randomly select a stub.
If it is a type-1 stub, we connect it to a randomly selected
type-1 stub in the associated block.  If it is a type-2 block,
we connect it to a randomly selected type-2 stub out of all
type-2 stubs.  We repeat this process until all stubs are
connected. To generate a generalized configuration model
with assortative mixing property, we select permutation
of blocks such that a block of stubs with large degrees is
associated with another block of large degrees.
To generate a network with disassortative
mixing, we select permutation such that a block of large
degrees is assciated with a block of small degrees.
We present the detail of the construction algorithm in
Section 2.
For this model, we derive a closed form for the Pearson correlation coefficient of
the degrees of two neighboring vertices.  From the Pearson
correlation coefficient of degrees we show that our constructed
network can be assortative or disassortative as desired.

In this paper we present an application of the proposed random graph model.
We consider a percolation analysis of the generalized configuration model.
Percolation has been a powerful tool to study
network resilience under breakdowns or attacks.  Cohen et. al [6]
studied the resilience of networks with scale-free degree distributions.
Particularly, Cohen et. al studied the stability of such networks,
including the Internet, subject to
random crashes.  Percolation has also been used to study disease
spreading in epidemic networks [5,16,25].
Percolation has been used to study the effectiveness of immunization
or quarantine to confine a disease.
Schwartz {\em et al.} [27] studied percolation
in a directed scale-free network.  Newman [19] and
V\'{a}zquez {\em et al.} [28] studied percolation in networks with
degree correlation.  V\'{a}zquez {\em et al.} assumed general
random networks, their solution involves with the eigenvalues of a $D\times D$
matrix, where $D$ is the total number of degrees in the network.
The percolation analysis of our model involves with solving roots of
a simultaneous system of $b$ nonlinear equations, where $b$ is the number of
blocks in the generalized configuration model. Since $b$ is typically a small
integer, we have significantly reduced the complexity.

The rest of this paper is organized as follows.
In Section 2 we present our construction method of a
random network.  In Section 3 we derive a closed form
for the joint degree distribution of two randomly selected neighboring vertices
from a network constructed by the algorithm in Section 2.
In Section 4, we show that the Pearson degree correlation
function of two neighboring vertices is linear.  We then show
how permutation function $h$ should be selected such that a
constructed random graph is associatively or disassortatively mixed.
In Section 5 we present a percolation analysis of this model.
Numerical examples and simulation results are presented
in Section 6.  Finally, we give conclusions in Section 7.

\section{Construction of a Random Network}

Research on random networks was pioneered by \er [8].
Although Erd\H{o}s-R\'{e}nyi's model allows researchers to study many
network problems, it is limited in that the vertex
degree has a Poisson distribution asymptotically as the network
grows in size.  The configuration model [2,15]
can be considered as an extension of the \er model that
allows general degree distributions.
Configuration models have been used successfully to study the size of
giant components.  It has been used to study network resilience when
vertices or edges are removed.  It has also been used to study
the epidemic spreading on networks.  We refer the readers to
[21] for more details.
In this paper we propose an extension of the
classical configuration model.
This model generates networks with specified degree sequences.
In addition, one can specify a positive or a
negative degree correlation
for the model. Let there be $n$ vertices and let $p_k$ be the
probability that a randomly selected vertex has degree $k$.
We sample the degree distribution $\{p_k\}$ $n$ times to obtain
a degree sequence $k_1, k_2, \ldots, k_n$ for the $n$ vertices.
We give each vertex $i$ a total of $k_i$ stubs.
There are $2m=\sum_{i=1}^n k_i$
stubs totally, where $m$ is the number of edges of the network.
In a classical configuration model, we randomly select an unconnected
stub, say $s$, and connect it to another randomly selected unconnected
stub in $[1, 2m]-\{s\}$.
We repeat this process until all stubs
are connected.  The resulting network can be viewed as a matching of
the $2m$ stubs.  Each possible matching occurs
with equal probability.  The consequence of this construction is
that the degree correlation of a randomly selected pair of neighboring
vertices is zero.
To achieve nonzero degree correlation, we arrange the $2m$
stubs in ascending order (descending order will
also work) according to the degree of the
vertices, to which the stubs belong.  We label the stubs accordingly.
We partition the $2m$ stubs into $b$ blocks evenly.  We select
integer $b$ such that $2m$ is divisible by $b$.
Each block has $2m/b$ stubs.
Block $i$, where $i=1, 2, \ldots, b$, contains stubs
$(i-1)(2m/b)+j$ for $j=1, 2, \ldots, 2m/b$.
Next, we choose a permutation
function $h$ of $\{1,2, \ldots, b\}$.
If $h(i)=j$, we say that block $j$ is associated with block $i$.
In this paper we select $h$ such that
\[
h(h(i))=i,
\]
{\em i.e.,} if blocks $i$ and $h(i)$ are mutually associated with
each other.
In each block, we randomly designate $\lceil 2mq/b\rceil$
stubs as type 1 stubs, where $q$ is a parameter in
the range $[0, 1)$.  Other stubs are designated as type 2 stubs.
Randomly select an unconnected stub.  Suppose that this stub
is in block $i$. If it is a type 1 stub, connect
this stub to a randomly selected unconnected type 1 stub in block $h(i)$.
If it is a type 2 stub, connect it to a randomly selected
unconnected type 2 stub in $[1, 2m]$. We repeat this process until
all stubs are connected.
The construction algorithm is shown in Algorithm 1.
\begin{algorithm}
\SetAlgoLined
{\bf Inputs}: degree sequence $\{k_i: i=1, 2, \ldots, n\}$\;
{\bf Outputs}: graph $(G,V,E)$\;
Create $2m$ stubs arranged in descending order\;
Divide $2m$ stubs into $b$ blocks evenly.  Initially,
all stubs are unconnected. For each block, randomly
designate $\lceil 2mq/b\rceil$ stubs as type 1 stubs.
All other stubs are designated as type 2 stubs\;
\While{there are unconnected stubs}{
Randomly select a stub. Assume that the stub is in block $i$\;
\eIf{type 1 stub}{
connect this stub with a randomly
selected unconnected type 1 stub in block $h(i)$\;
}{
connect this stub with a randomly
selected unconnected type 2 stub in $[1, 2m]$\;
}}
\caption{Construction Algorithm}\label{alg1}
\end{algorithm}

We make a few remarks.  First, note that in
networks constructed by this algorithm, there are $mq$ edges that have
two type-1 stubs on their two sides.  These edges
create degree correlation in the network.
On the other hand, there are $m(1-q)$
edges in the network that have two type-2
stubs on their two sides.  These edges do not contribute
towards degree correlation in the network.
Second, random networks constructed by this algorithm possess
the following property.  A randomly selected stub
connects to another randomly selected in the associated
block with probability $q$.  With probability $1-q$, this stub
connects to a randomly selected stub in $[1, 2m]$.
Finally, note that standard
configuration models can have multiple edges connecting two particular
vertices.  There can also be edges connecting a vertex to itself.
These are called multiedges and self edges.  In our constructed
networks, multiedges and self edges can also exist.
However, it is not difficult to show that the expected density
of multiedges and self edges approaches to zero as $n$
becomes large.  Due to space limit, we shall not address this
issue in this paper.

\section{Joint Distribution of Degrees}

Consider a randomly selected
edge in a random network constructed by the algorithm
described in Section 2.  In this section we analyze the joint
degree distribution of the two vertices at the two ends
of the edge.

We randomly select a vertex and let $Z$ be the degree of this vertex.
Since the selection of vertices is random,
\[
\Pr(Z=k_i)=\frac{1}{n}
\]
for $i=1, 2, \ldots, n$.
Thus, the expectation of $Z$ is
\[
\ex(Z)=\frac{\sum_{i=1}^n k_i}{n}= \frac{2m}{n}.
\]
The expectation of $Z$ can also be expressed as
\beq{extildeX}
\ex(Z)=\sum_{k=0}^\infty k p_k.
\eeq
The expected number of stubs of the network is $\ex(Z)\cdot n$.
We would like to evenly allocate these stubs into $b$ blocks such that each block has
$n\ex(Z)/b$ stubs on average.  We make the following assumption.
\bassum{1}
The degree distribution $\{p_k\}$ is said to satisfy this assumption if
one can find mutually disjoint sets $H_1, H_2, \ldots, H_{b}$, such
that
\[
\bigcup_{i=1}^b H_i=\{0, 1, 2, \ldots\}
\]
and
\beq{evenblock}
\sum_{k\in H_{i}} k p_k =\ex(Z)/b
\eeq
for all $i=1, 2, \ldots, b$.  In addition, we assume that the degree sequence
$k_1, k_2, \ldots, k_n$ sampled from the distribution $\{p_k\}$ can be evenly
placed in $b$ blocks.
Specifically, there exist mutually disjoint sets $H_1, H_2, \ldots, H_{b}$
that satisfy
\begin{enumerate}
\item $\bigcup_{i=1}^b H_i=\{1, 2, \ldots, n\}$,
\item $k_i\ne k_j$ for any $i\in H_{\ell_1}$,
$j\in H_{\ell_2}$, $\ell_1\ne\ell_2$, and
\item $\sum_{j\in H_i} k_j=2m/b$ for all $i=1, 2, \ldots, b$.
\end{enumerate}

\eassum
We now present some remarks on Assumption 1.

{\bf Remarks.}
\begin{itemize}
\item Note that the construction algorithm
described in Section 2
works without Assumption 1.  However, this assumption
allows us to derive a very simple expression for the joint
probability mass function (pmf) of $X$ and $Y$.  This
simple expression allows us to analyze
the assortativity and disassortativity of the model.
For degree sequences that do not satisfy
Assumption 1, the analyses in Sections 3, 4 and 5 are only
approximate.  In Section 6, we shall compare simulation results
of models constructed without Assumption 1 with analytical results.
\item Assumption 1 is quite restrictive.
In Section 6 we discuss how one modifies a degree distribution
to make the distribution satisfy Assumption 1.
We also remark that from $(2)$ one can view
\[
\tilde p_k=\frac{k p_k}{\ex(Z)}
\]
as a probability mass function.  Eq. (3) can be equivalently be
expressed as
\[
\sum_{k\in H_i} \tilde p_k=1/b
\]
for all $i=1, 2, \ldots, b$.  We can equivalently say that distribution
$\{\tilde p_k\}$ satisfies Assumption 1.
\item Finally, we remark that a  common
way to generate stubs from a degree distribution is to first generate
a sequence of uniform pseudo random variables over $[0, 1]$.
Then, transform the uniform random variables using
the inverse cumulative distribution function of the degree distribution [4].
This approach would encounter difficulties as far as Assumption 1 is concerned,
because the stubs produced
are not likely to be evenly allocated among blocks.  If the network is large, the
following approach based on proportionality can be used.  Specifically,
for degree $k$ with probability mass $p_k$, create $np_k$ vertices and
$nk p_k$ corresponding stubs.  If $n$ is large, the strong law of large numbers
ensures that this approach and the inversion
method produce approximately the same number of stubs.
Using this approach, the probability masses of the degree distribution
and the stubs sampled from the degree distribution
both satisfy Assumption 1 and can be placed
evenly in blocks at the same time.
\end{itemize}


We randomly select a stub in the range $[1, 2m]$.  Denote this stub
by $t$.  Let $v$ be the vertex, with which stub $t$ is associated.  Let $Y$ be
the degree of $v$.   Now connect
stub $t$ to a randomly selected stub according to the construction algorithm
in Section 2.  Let this stub be denoted by $s$.  Let $u$ be
the vertex, with which $s$ is associated, and let $X$ be the degree of $u$.
Since stub $t$ is randomly selected from range $[1, 2m]$, the distribution of $Y$ is
\beq{distY}
\Pr(Y=k)=\frac{n k p_k}{2m}
= \frac{k p_k}{\ex(Z)},
\eeq
where $Z$ is the degree of a randomly selected vertex.

To study the joint pmf of $X$ and $Y$,
we first study the conditional pmf of $X$, given $Y$, and the marginal pmf $X$.
In the rest of this section, we assume that Assumption 1 holds.
Suppose $x$ is a degree in set $H_i$.
The total number of stubs which are associated with
vertices with degree $x$ is $n x p_x$.  By Assumption 1,
all $nx p_x$ stubs are in block $i$.  We consider two cases, in which
stub $t$ connects to stub $s$.  In the first case, stub $t$ is of type 1.  This
occurs with probability $q$.  In this case, stub $t$ must belong to a vertex
with a degree in block $h(i)$.  With probability
\beq{q}
\frac{qnxp_x}{2mq/b-\delta(i, h(i))},
\eeq
the  construction algorithm in Section 2 connects $t$ to stub $s$,
where $\delta()$ is the delta function.  In the second case, stub $t$ is of type 2.
This occurs with probability $1-q$.  In this case, stub $t$ can be associated with a degree
in any block.
With probability
\beq{1-q}
\frac{(1-q)nx p_x}{2m(1-q)-1}
\eeq
the construction algorithm connects stub $t$ to stub $s$.
Combining the two cases in (5) and (6), we have
\beq{condp1}
\Pr(X=x | Y=y)
=\frac{q^2 nxp_x}{2mq/b-\delta(i, h(i))}+
\frac{(1-q)^2 nx p_x}{2m(1-q)-1},
\eeq
for $y\in H_{h(i)}$.
If $y\in H_j$ for $j\ne h(i)$,
\beq{condp2}
\Pr(X=x | Y=y)= \frac{(1-q)^2 nxp_x}{2m(1-q)-1}.
\eeq
Now assume that the network is large.  That is, we
consider a sequence of constructed graphs, in which
$n\to\infty$, $m\to\infty$, while keeping
$2m/n=\ex(Z)$.
Under this asymptotics, Eqs. (7) and (8)
converge to
\beq{condp-limit}
\Pr(X=x | Y=y)\to
\left\{\begin{array}{ll}
\frac{qb+(1-q)}{\ex(Z)}xp_x, & \quad y\in H_{h(i)} \\
\frac{1-q}{\ex(Z)}xp_x, & \quad y\in H_j, j\ne h(i).
\end{array}\right.
\eeq
From the law of total probability we have
\bear{marginal-eq}
\Pr(X=x)&=&\sum_{y\in H_{h(i)}} \Pr(X=x | Y=y)\Pr(Y=y)\nonumber\\
&&\quad+\sum_{j\ne h(i)}\sum_{y\in H_j}
 \Pr(X=x | Y=y)\Pr(Y=y).
\eear
Substituting (4) and (9) into
(10), we have
\beq{mpmf}
\Pr(X=x)
=\sum_{y\in H_{h(i)}} \frac{qb+(1-q)}{\ex(Z)}xp_x
\frac{y p_y}{\ex(Z)}
+\sum_{j\ne h(i)}\sum_{y\in H_j}
\frac{1-q}{\ex(Z)}xp_x\frac{y p_y}{\ex(Z)}.
\eeq
Since the partition of stubs is uniform,
\[
\sum_{y\in H_j} n y p_y=2m/b
\]
and thus,
\[
\sum_{y\in H_j} y p_y=\ex(Z)/b
\]
for any $j=1, 2, \ldots, b$.
Substituting this into (11), we have
\beq{mpmf2}
\Pr(X=x)=\frac{x p_x}{\ex(Z)}.
\eeq

From (9) we derive the joint pmf of $X$ and $Y$
\bear{joint}
&&\Pr(X=x, Y=y)
=\Pr(X=x | Y=y)\Pr(Y=y)\nonumber\\
&&=\left\{\begin{array}{ll}
\left(b q+1-q\right)\frac{x p_x}{\ex(Z)}
\frac{y p_y}{\ex(Z)},
& \quad y\in H_{h(i)}, x\in H_i\\
(1-q)\frac{x p_x}{\ex(Z)}\frac{yp_y}{\ex(Z)}, &
\quad x\in H_i, y\in H_j,  j\ne h(i)
\end{array}\right. \nonumber\\
&&=C_{ij}\frac{xy p_x p_y}{(\ex(Z))^2},
\eear
where
\beq{C}
C_{ij}=\left\{\begin{array}{ll}b q+1-q, & \quad h(i)=j\\
1-q, & \quad h(i)\ne j.\end{array}\right.
\eeq

We summarize the results in the following theorem.
\bthe{pmf}
Let $\G$ be a graph generated by the construction algorithm
described in Section 2 based on a sequence of
degrees $k_1, k_2, \ldots, k_n$.  Randomly select an
edge from $\G$.  Let $X$ and $Y$ be the degrees of the two
vertices at the two ends of the edge.  Then, the marginal
pmf of $X$ and $Y$ are given in (12) and (4),
respectively.  The joint pmf of $X$ and $Y$ is given
in (13).
\ethe


\section{Assortativity and Disassortativity}

In this section, we present an analysis of the Pearson degree correlation
function of two random neighboring vertices.
The goal is to search for permutation function $h$ such that
the numerator of (1) is non-negative (resp. non-positive)
for the network constructed in this section.

From (12), we obtain the expected value of $X$
\beq{exp-X}
\ex(X)=\sum_x x \Pr(X=x)
=\sum_{i=1}^b \sum_{x\in H_i}\frac{x^2 p_x}{\ex(Z)}
=\frac{1}{\ex(Z)}\sum_{i=1}^b u_i,
\eeq
where
\beq{def-u}
u_i \defeq \sum_{x\in H_i} x^2 p_x.
\eeq
Now we consider the expected value of the product $XY$.
We have from (13) that
\bear{prod1}
&&\ex(XY)=\sum_{x}\sum_y xy\Pr(X=x, Y=y)
=\sum_{i=1}^b\sum_{j=1}^b \sum_{x\in H_i} \sum_{y\in H_j}
\frac{C_{ij} x^2 y^2 p_x p_y}{(\ex(Z))^2}\nonumber \\
&&=\sum_i\sum_j \frac{C_{ij}u_i u_j}{(\ex(Z))^2}
=\frac{1}{(\ex(Z))^2}\left((1-q)\sum_i\sum_j u_i u_j
+ qb \sum_i u_i u_{h(i)}\right).
\eear
Note from (15) and (17) that
\beq{numerator1}
\ex(XY)-\ex(X)\ex(Y)
=\frac{q}{(\ex(Z))^2}\Big (b\sum_i u_i u_{h(i)}-
\sum_i\sum_j u_i u_j \Big ).
\eeq
Based on (18), we summarize the Pearson degree
correlation function in the following theorem.
\bthe{Pearson}
Let $\G$ be a graph generated by the construction algorithm
in Section 2.  Randomly select an edge from the
graph.  Let $X$ and $Y$ be the degrees of the two vertices
at the two ends of this edge.  Then, the Pearson degree
correlation function of $X$ and $Y$ is
\beq{pdcf}
\rho(X, Y)=cq,
\eeq
where
\[
c=\frac{b\sum_i u_i u_{h(i)}-
\sum_i\sum_j u_i u_j}{\sigma_X\sigma_Y(\ex(Z))^2},
\]
and $\sigma_X$ and $\sigma_Y$ are the standard deviation
of the pmfs in (12) and (4).
\ethe

In view of (19), the sign of $\rho(X, Y)$ depends on
the constant $c$.
To generate assortative (resp. disassortative) mixing random graphs
we sort $u_i$'s in descending order first and then choose the
permutation $h$ that maps the largest number of $u_i$'s
to the largest (resp. smallest) number of $u_i$'s.
This is formally stated in the following corollary.
\bcor{main} Let $\pi(\cdot)$ be the permutation such that
$u_{\pi(i)}$ is the $i^{th}$ largest number
among $u_i$, $i=1,2, \ldots, b$, i.e.,
$$u_{\pi(1)} \ge u_{\pi(2)} \ge \ldots \ge u_{\pi(b)}.$$
\begin{description}
\item[(i)] If we choose the permutation $h$ with
$h(\pi(i))=\pi(i)$ for all $i$, then the constructed random graph is
assortative mixing.
\item[(ii)] If we choose the permutation $h$ with
$h(\pi(i))=\pi(b+1-i)$ for all $i$, then the constructed random graph is
disassortative mixing.
\end{description}
\ecor
The proof of Corollary 1 is based on the famous Hardy,
Littlewood and P\'olya rearrangement inequality
(see e.g., the book [13], pp. 141).
\bprop{HLP} {\bf (Hardy, Littlewood and P\'olya rearrangement inequality)} If $u_i, v_i$, $i=1,2, \ldots, b$ are two sets of real number.
Let $u_{[i]}$ (resp. $v_{[i]}$) be the $i^{th}$ largest number
among $u_i$, $i=1,2, \ldots, b$  (resp. $v_i$, $i=1,2, \ldots, b$).
Then
\beq{HLP1111}
\sum_{i=1}^b u_{[i]} v_{[b-i+1]} \le \sum_{i=1}^b u_i v_i \le \sum_{i=1}^b u_{[i]} v_{[i]}.
\eeq
\eprop

\bproof (Corollary 1)
\noindent (i) Consider the circular shift permutation $\sigma_j(\cdot)$
with $\sigma_j(i)=(i+j-1\;\mbox{mod}\;b)+1$ for $j=1,2, \ldots, b$.
From symmetry, we have $\sigma_j(i)=\sigma_i(j)$.
Thus,
\beq{main1111}
\sum_{i=1}^b \sum_{j=1}^b u_i u_j=\sum_{i=1}^b \sum_{j=1}^b u_i u_{\sigma_i(j)}=
\sum_{j=1}^b \sum_{i=1}^b u_i u_{\sigma_j(i)}.
\eeq
Using the upper bound of the Hardy, Littlewood and P\'olya
rearrangement inequality in (21) and $h(\pi(i))=\pi(i)$ yields
\beq{main2222}
\sum_{i=1}^b u_i u_{\sigma_j(i)} \le \sum_{i=1}^b u_{[i]} u_{[i]}
=  \sum_{i=1}^b u_{\pi(i)} u_{h(\pi(i))} = \sum_{i=1}^b u_i u_{h(i)}.
\eeq
In view of (18) and (21), we conclude that
the generated random graph is assortative mixing.

\noindent (ii)
Using the lower bound of the Hardy, Littlewood and P\'olya rearrangement inequality in (21) and $h(\pi(i))=\pi(b+1-i)$ yields
\beq{main2222b}
\sum_{i=1}^b  u_i u_{\sigma_j(i)}\ge \sum_{i=1}^b u_{[i]} u_{[b+1-i]}
=\sum_{i=1}^b u_{\pi(i)} u_{h(\pi(i))} = \sum_{i=1}^b u_i u_{h(i)}.
\eeq
In view of (18) and (21), we conclude that
the generated random graph is disassortative mixing.
\eproof

\section{An Application: Percolation}

In this section we present a percolation analysis of the generalized configuration
model.

We consider node percolation of a random network with $n$ vertices.
Recall that we define $Z$ to be the degree of a randomly selected
vertex in the network.  Let $p_k=\Pr(Z=k)$ be given and let $\ex(Z)$ be
the expected value of $Z$.

Let $\phi$ be the probability that a node stays
in the network after the percolation.  That is, $1-\phi$ is the probability that
a node is removed from the network.  In the literature of percolation
analysis, $\phi$ is called the occupation probability.
We assume that $\phi\in (0,1)$.
Let $\alpha_i$ be the probability that along
an edge with one end attached to a stub in block $i$, one can {\em not}
reach a giant component.   Let $\eta_i$ be the probability that a randomly selected
vertex from block $i$ is in a giant component after the
random removal of vertices.  Then,
\beq{eta}
\eta_i=\phi\sum_{k\in H_i} p_k \left(1-\alpha_i^k\right).
\eeq
Let $\eta$ be the probability that a randomly selected
vertex is in a giant component after the
random removal of vertices.  Then,
\beq{sgc}
\eta=\sum_{i=1}^b \eta_i \sum_{k\in H_i} p_k.
\eeq
We now derive a set of equations for $\alpha_i$, $i=1, 2, \ldots, b$.
We randomly select an edge.  Call this edge $e$.
Let $D$ be the event that $e$ does not
connect to a giant component.  Let $B_i$ be the event
that one end of this edge is associated with a stub in block $i$.
Suppose that the other end of $e$ is attached to a vertex called
$v$.
Then by the law of total probability we have
\beq{lawtp}
\Pr(D | B_i)=\sum_{j=1}^b\sum_{k=1}^\infty \Pr(D | Y=k, B_j, B_i)
\Pr(Y=k, B_j | B_i),
\eeq
where $Y$ is the degree of $v$ and $B_j$ is the event that
vertex $v$ is in block $j$.
According to (9), we have
\beq{condp3}
\Pr(Y=k, B_j | B_i)=\left\{\begin{array}{ll}
\frac{qb+(1-q)}{\ex(Z)}k p_k, & \quad k\in H_{h(i)}, \\
\frac{1-q}{\ex(Z)}k p_k, & \quad k\in H_j, j\ne h(i).
\end{array}\right.
\eeq
If vertex $v$ is removed from the network through percolation, then
edge $e$ does not lead to a giant component.  This occurs with probability
$1-\phi$.  With probability $\phi$, vertex $v$ is not removed.  Conditioning
on $Y=k$, edge $e$ does not lead to a giant component if all the $k-1$
edges of $v$ do not.  In addition, conditioning on $B_j$, event $D$ is
independent from event $B_i$.  Combining these facts together, we have
\bear{condp4}
\Pr(D | Y=k, B_j, B_i)&=& \Pr(D | Y=k, B_j) \nonumber\\
&=& 1-\phi+\phi \alpha_j^{k-1}.
\eear
Substituting (27) and (28) into (26), we have

\bear{recu1}
\alpha_i &=&\sum_{k\in H_{h(i)}}\left(1-\phi+\phi \alpha_{h(i)}^{k-1}\right)
\frac{(bq+1-q)k p_k}{\ex(Z)}\nonumber\\
&&\quad+\sum_{j=1,j\ne h(i)}^b \sum_{k\in H_j}\left(1-\phi+\phi \alpha_j^{k-1}\right)
\frac{(1-q)k p_k}{\ex(Z)}.
\eear
Let
\beq{defg}
g_i(x)=\sum_{k\in H_i}\frac{k p_k x^{k-1}}{\ex(Z)}
\eeq
for $i=1, 2, \ldots, b$.
Combining constant terms, we rewrite (29) in terms of $g_i(z)$, {\em i.e.}
\beq{recu2}
\alpha_i = 1-\phi+\phi\left((bq+1-q)g_{h(i)}(\alpha_{h(i)})
+(1-q)\sum_{j=1,j\ne h(i)}^b g_j(\alpha_j)\right).
\eeq
Expressing (31) in the form of vectors, we have
\beq{system1}
\bfalpha=\bff(\bfalpha),
\eeq
where
$\bfalpha$ is a vector in $[0,1]^b$ and $\bff$ is a vector function that maps from
$[0,1]^b$ to $[0,1]^b$.  In this section, we use boldface letters to denote
vectors.  
The $i$-th entry of $\bff(\bfalpha)$ is denoted by
\beq{deff}
f_i(\bfalpha)= 1-\phi+\phi\left((bq+1-q)g_{h(i)}(\alpha_{h(i)})
+(1-q)\sum_{j=1, j\ne h(i)}^b g_j(\alpha_j)\right).
\eeq
Solutions of (32) are called the fixed points of the function $\bff$.

Note that $\alpha_i=1$ for all $i=1, 2, \ldots, b$, is always a root
of (32).
Denote point $(1, 1, \ldots, 1)$ by $\bfone$.
We are searching for a condition under which $\bfalpha=\bfone$ is the only
solution of (32) in $[0, 1]^b$, and a condition under which (32)
has additional solutions.
Define
\beq{Jm-def}
\bfJ(\bfa)=\left.\left(\begin{array}{cccc}
\frac{\partial f_1(\bfsx)}{\partial x_1} & \frac{\partial f_1(\bfsx)}{\partial x_2}
& \ldots & \frac{\partial f_1(\bfsx)}{\partial x_b}  \\
\frac{\partial f_2(\bfsx)}{\partial x_1} & \frac{\partial f_2(\bfsx)}{\partial x_2}
& \ldots & \frac{\partial f_2(\bfsx)}{\partial x_b}  \\
\vdots & \vdots & \ddots & \vdots \\
\frac{\partial f_b(\bfsx)}{\partial x_1} & \frac{\partial f_b(\bfsx)}{\partial x_2}
& \ldots & \frac{\partial f_b(\bfsx)}{\partial x_b} \end{array}\right)\right|_{\bfsx=\bfsa},
\eeq
where $\bfa=(a_1, a_2, \ldots, a_b)$ is a point in $[0, 1]^b$.
Matrix $\bfJ(\bfa)$ is called the Jacobian matrix of function $\bff(\bfx)$.
For function $\bff$ defined in (33), the Jacobian matrix has the following
form
\beq{Jm}
\bfJ(\bfa)=\phi (b q \bfH+(1-q)\bfone_{b\times b}) 
\bfD\{g_1'(a_1), g_2'(a_2), \ldots, g_b'(a_b)\},
\eeq
where $\bfone_{b\times b}$ is a $b\times b$ matrix of unities,
and $\bfD\{g_1'(a_1), g_2'(a_2), \ldots, g_b'(a_b)\}$ is a diagonal matrix.
In (35), matrix $\bfH$ is a permutation matrix whose $(i, j)$ entry
is one if $j=h(i)$, and is zero otherwise.
Let $\phi\lambda_1, \phi\lambda_2, \ldots, \phi\lambda_b$ be the eigenvalues of
$\bfJ(\bfone)$ with
\[
|\lambda_1|\ge |\lambda_2|\ge \ldots\ge |\lambda_b|.
\]
Since $g_j$ is a power series with non-negative coefficients for all
$j$, $g_j'$ is strictly increasing and $g_j'(1)>0$.  Thus, $\bfJ(\bfone)$ is a positive
matrix.  According to the Perron-Frobenius theorem [14,11],
$\phi\lambda_1$ is real, positive and strictly larger than $\phi\lambda_2$
in absolute value.  In addition, there exists an eigenvector $\bfv$
associated with the
dominant eigenvalue that is positive component-wise.

The existence of roots of (32) is summarized in the
following main result.
\bthe{the1}
Let
\beq{phi-c}
\phi^\star=1/\lambda_1.
\eeq
The solution of (32) can be in one of two cases.
\begin{enumerate}
\item If $0< \phi< \phi^\star$, point $\bfone$ is an attracting
fixed point.   In addition, it is the only fixed point in $[0, 1]^b$.
\item If $\phi^\star < \phi< 1$,
point $\bfone$ is either a repelling fixed point or a saddle point
of the function $\bff$ in (32).  There exists
another fixed point in $[0,1)^b$.
This additional fixed point is an attracting fixed point.
\end{enumerate}
\ethe

The proof of Theorem 3 is presented in the appendix.
Note that in case 1 of Theorem 3, the only root is $\bfalpha=\bfone$.
From (24), $\eta_i=0$ for all $i=1, 2, \ldots, b$.  It follows
that $\eta=0$ and the network has no giant component.
In case 2, the network has a giant component whose size is determined by the additional
fixed point.

We first study the behavior of $\bff$ in the neighborhood of $\bfone$.
We consider the following iteration
\beq{iteration}
\bfx_{n+1}=\bff(\bfx_n),\qquad n=0,1, 2, \ldots
\eeq
where the initial vector $\bfx_0$ is in the neighborhood of the fixed point $\bfone$.
Assume that $g_i(x)$ can be linearized,
i.e. $g_i(x)$ can be approximated by keeping two terms in its Taylor
expansion around one
\beq{Taylor}
g_i(x)\approx g_i(1)+g_i'(1)(x-1)
\eeq
for all $i=1, 2, \ldots, b$.  Now substituting (38) into
(37) and noting that
\begin{eqnarray*}
&&g_i(1) =1/b \\
&& (bq+1-q)g_{h(i)}(1)+(1-q)\sum_{j=1, j\ne h(i)}^b g_j(1)=1.
\end{eqnarray*}
for all $i=1, 2, \ldots, b$, we obtain the following matrix equation
\beq{iteration2}
\bfx_{n+1}-\bfone = \bfJ(\bfone) (\bfx_n -\bfone),
\eeq
where we recall that $\bfJ(\bfone)$ is the Jacobian matrix
stated in (35).
Substituting (39) repeatedly into itself, we obtain
\[
\bfx_{n}-\bfone=(\bfJ(\bfone))^{n} (\bfx_0-\bfone).
\]
If the dominant eigenvalue $\phi\lambda_1<1$,
$\bfx_n-\bfone\to\bfzero$ and $\bfone$ is an attracting fixed point.
If all eigenvalues are greater than one in absolute value,
$\bfx$ moves away from $\bfone$.  In this case, $\bfone$
is a repelling fixed point.  Suppose that some eigenvalues are greater than one
and some are less than one in absolute values.
In this case, point $\bfone$ is called a saddle point.
Point $\bfx_n$ is attracted
to $\bfone$, if $\bfx_0-\bfone$ is a linear combination of the eigenvectors
associated with eigenvalues smaller than one in absolute values.
Otherwise, $\bfx_n$ moves away from $\bfone$.

\section{Numerical and Simulation Results}

We report our simulation results in this section.
Recall that we derive the degree covariance of two neighboring
vertices based on Assumption 1.  Assumption 1 is somewhat
restrictive.  For degree sequences that do not satisfy
Assumption 1, the analyses in Sections 3 and 4 are only
approximate.  In this section, we compare simulation results
with the analytical results in Section 4.

We have simulated
the construction of networks with 4000 vertices.  We use the batch mean
simulation method to control the simulation variance.  Specifically,
each simulation is repeated 100 times to obtain 100 graphs.
Eq. (1) was applied to compute the assortativity coefficient
for each graph.   One average is computed for every twenty repetitions.
Ninety percent confidence intervals are computed based on five averages.
We have done extensive number of simulations for uniform and Poisson
distributed degree distributions.
We have found that simulation results on Pearson degree
correlation coefficient agree extremely well with (19)
for a wide range of $b$ and $q$.  Due to space limit, we do not
present these results in the paper.
We have also simulated power-law degree distributions. Specifically,
we assume that the exponent of the power-law distribution
is negative two, {\em i.e.}, $p_k\approx k^{-2}$ for large $k$.
We first fix $b$ at six.
The degree correlations for power-law degree distributions are
shown in Figure 1 and Figure 2 for
an assortatively mixed network and a disassortatively mixed
network, respectively.  The discrepancy between the simulation
result and the analytical result is quite noticeable
in Figure 1 when $q$ is large, while the two results agree very well
in Figure 2.  This is because power-law distributions can
generate very large sample values for degrees.  As a result,
Assumption 1 may fail in this case.  We decrease $b$ to two,
which increases the block size.  The corresponding Pearson
degree correlation function for an disassortatively mixed network
is presented in Figure 3.  One can see that the
approximation accuracy is dramatically increased as the block
size is increased.

\bfig{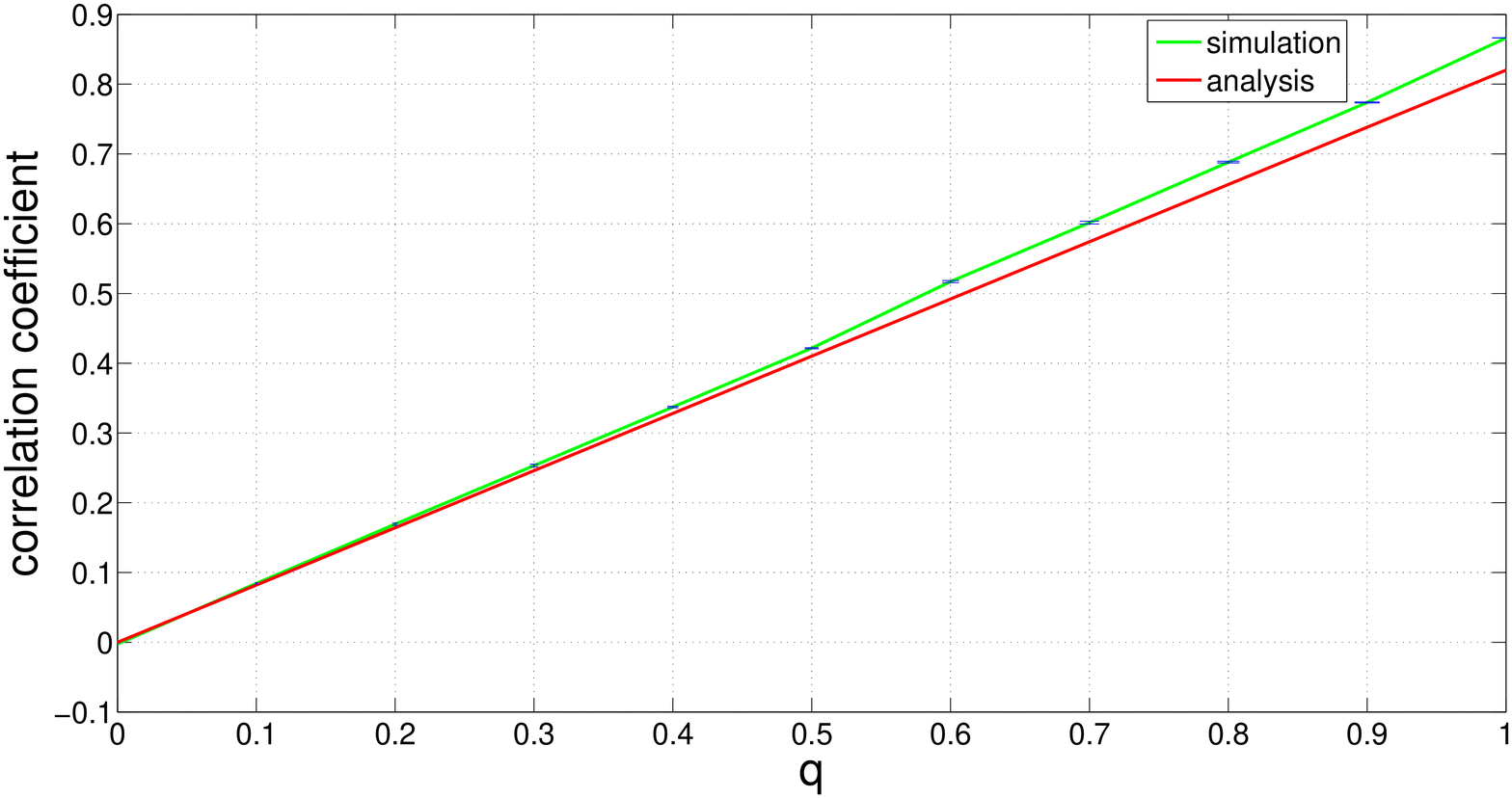}{4.75in}
\efig{powerlaw-p6}{Degree correlation of an assortative model.
Power-law degree distribution and $b=6$}
\bfig{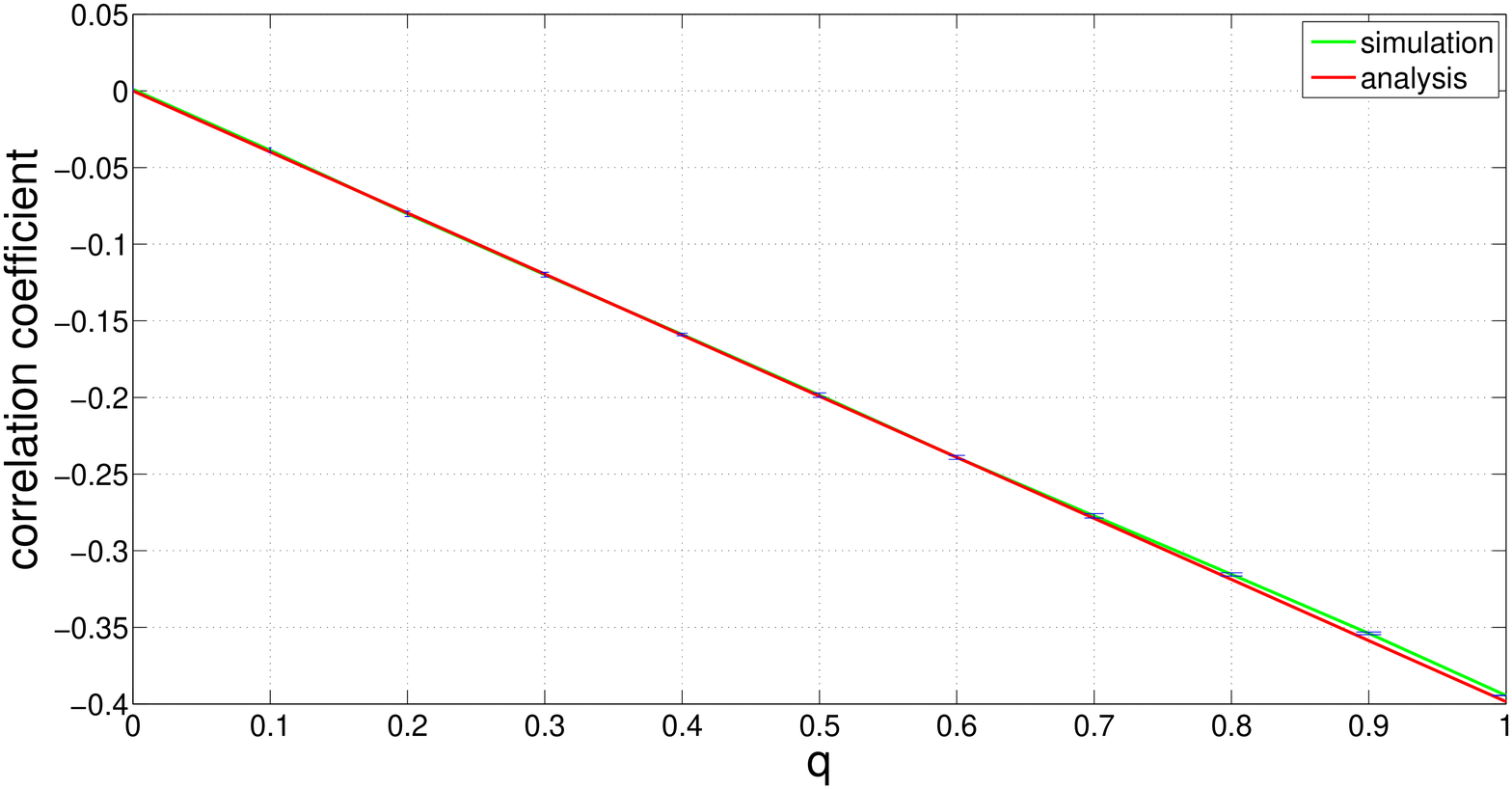}{4.75in}
\efig{powerlaw-n6}{Degree correlation of a disassortative model.
Power-law degree distribution and $b=6$}
\bfig{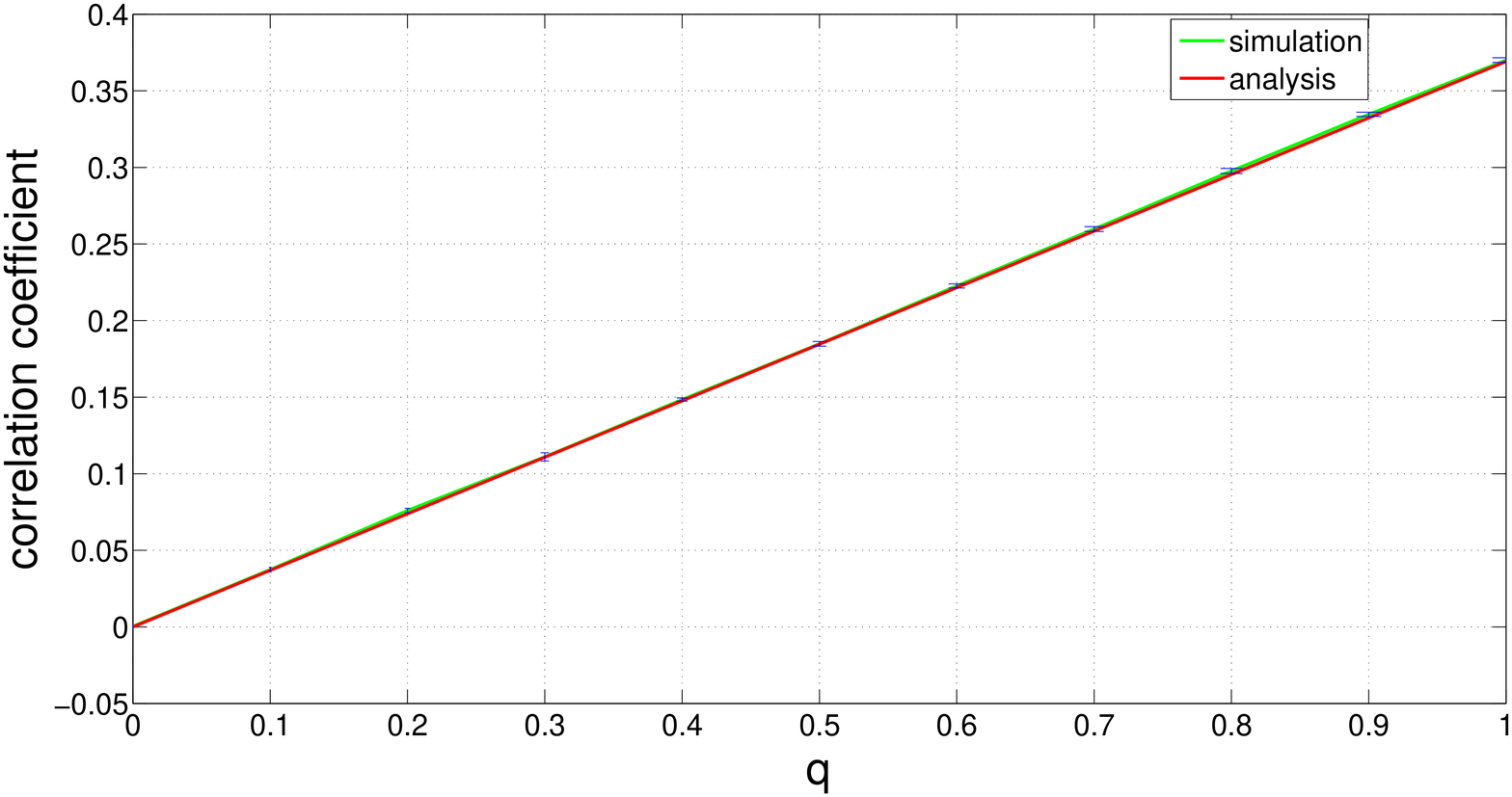}{4.75in}
\efig{powerlaw-p2}{Degree correlation of a assortative model.
Power-law degree distribution and $b=2$}

For percolation analysis, we study the critical value of $\phi$.
We assume that degrees are geometrically distributed.
However, geometrical distributions do not satisfy Assumption 1.  Assumption 1 is essential.
Without this assumption, $\bfone$ is not a fixed point and numerical calculations would fail.
We can adjust the probability masses to make Assumption 1 hold.
We illustrate this modification for the $b=2$ case.
We start with a geometric degree distribution $(1-p)p^k$, where $k=0, 1, \ldots$, and $p=2/3$.
The corresponding $\ex(Z)=2$.  We thus have
\[
\tilde p_k=k (1-p)p^k/2,\quad k=0, 1, \ldots
\]
We move part of the probability mass from $\tilde p_4$ to $\tilde p_5$.
After this modification, the distribution $\{\tilde p_k\}$ becomes
\beq{tp-example}
\tilde p_k=\left\{\begin{array}{ll}
k(1-p)p^k/2 & \mbox{$k\ge 0$, $k\ne 4$, $k\ne 5$;}\\
2(1-p)p^4-0.0782 & \mbox{if $k=4$;}\\
5(1-p)p^5/2 +0.0782 & \mbox{if $k=5$}.\end{array}\right.
\eeq
Let $H_1=\{0, 1, 2, 3, 4\}$ and $H_2=\{k: k \ge 5\}$.  It is easy to verify that
$\{\tilde p_k\}$ satisfies Assumption 1.

We study $b=2$ and $b=3$.  In both cases, we study two permutations
of blocks suggested in Section 4 for assortativity and disassortativity.
For assortative networks, $h(i)=i$.  For disassortative networks,
$h(i)=b+1-i$. In the case of $b=3$, we have also studied a rotational
permutation, i.e., $h(i)=((i+1) \mod b)+1$.  The critical values of $\phi$ obtained
using (36) are shown in Table 1.  We also numerically
calculate the critical values of $\phi$.  In this numerical study, we
gradually decrease $\phi$ until (32) fails to have a solution
in the interior of $[0, 1)^b$.  From these results, we see that the critical
values of $\phi$ obtained from (36) agree very well with those obtained
numerically.

Finally, we study the giant component sizes of the generalized configuration models.
We numerically solve (32) to obtain vector $\bfalpha$, and then compute
$\eta$ using (25).  In this study, we continue to assume that degrees are
geometrically distributed as we did in the study of Table 1.  The giant component
sizes are shown in Figure 4.  From this figure, we see that assortative networks
have smaller percolation thresholds than disassortative networks.  Hence, giant
components emerge more easily in assortative networks.  However,
disassortative networks tend to have larger giant component sizes than
assortative networks for large $\phi$.
The effect of assortativity and disassortativity to the giant component
sizes and the percolation thresholds observed in this example agrees with
that observed in Newman [19].
For the effect of $q$, larger values of $q$ decrease the percolation thresholds
and the giant component sizes of assortative networks.  On the other hand,
larger values of $q$ increase the percolation thresholds and the giant component sizes of disassortative networks.

\begin{table*}[!t]
\begin{center}
\begin{tabular}{|l|l|r|r|r|} \hline
 & & $q=0.2$ & $q=0.5$ & $q=0.8$ \\ \hline
\multirow{3}{*}{$b=2$, assortativity} 
& $\phi^\star$ & 0.22662 & 0.19518 & 0.16692 \\ \cline{2-5}
   & numerical  & 0.22662 & 0.19513 & 0.16688 \\ \hline
\multirow{3}{*}{$b=2$, disassortativity} 
& $\phi^\star$ & 0.26715 & 0.29237 & 0.31231 \\ \cline{2-5}
   &numerical  & 0.26711 & 0.29237 & 0.31229 \\ \hline
\multirow{3}{*}{$b=3$, assortativity} 
& $\phi^\star$ & 0.22252 & 0.18095  & 0.14540 \\ \cline{2-5}
  & numerical  & 0.22251 & 0.18092 & 0.14537 \\ \hline
\multirow{3}{*}{$b=3$, disassortativity}  
& $\phi^\star$ & 0.27442 & 0.30784 & 0.32967 \\ \cline{2-5}
  & numerical  & 0.27438 & 0.30782 &  0.32965 \\ \hline
\multirow{3}{*}{$b=3$, rotator}  
& $\phi^\star$ & 0.26572 & 0.29682 & 0.33182  \\ \cline{2-5}
  & numerical  & 0.26571 & 0.29682 & 0.33181  \\ \hline
 \end{tabular}\end{center}
\caption{Critical values of $\phi$. \label{cv}}
\end{table*}

\bfig{size-giant-component}{4.75in}
\efig{fgc}{Size of the giant component vs. $\phi$.}

\section{Conclusions}

In this paper we have presented an extension of the classical configuration
model.  Like a classical configuration model, the extended
configuration model allows users to specify an arbitrary degree
distribution.  In addition,
the model allows users to specify a positive or a negative
assortative coefficient.  We derived a closed form for the assortative
coefficient of this model.  We verified our result with simulations.

\begin{acknowledgement}
This
research was supported in part by the Ministry of Science and Technology,
Taiwan, R.O.C., under Contract 105-2221-E-007-036-MY3.
\end{acknowledgement}

\appendix {\bf Appendix}

In this appendix we prove Theorem 3.  To achieve this,
we need a matrix version of
the mean value theorem.  We state the result in the following lemma.
\blem{mvt}
Suppose that $\bfx$ and $\bfy$ are two points in $[0, 1]^b$.
Then, there exists constants $c_i$ in the open intervals
$(\min(x_i, y_i), \max(x_i, y_i))$, such that
\beq{vector-mvt}
\bff(\bfx)-\bff(\bfy)=\bfJ(\bfc)(\bfx - \bfy),
\eeq
where $\bfc=(c_1, c_2, \ldots, c_b)$.
\elem
\begin{proof}[Lemma 1]
Suppose that $\bfx$ and $\bfy$ are
two points in $[0, 1]^b$.  Consider
\bear{diff-f}
f_i(\bfx)-f_i(\bfy)&=&\phi(bq+1-q)\left(g_{h(i)}(x_{h(i)})-g_{h(i)}(y_{h(i)})\right) \nonumber\\
&&\ +\phi(1-q)\sum_{j=1, j\ne h(i)}^b \left(g_j(x_j)-g_j(y_j)\right).
\eear
Since function $g_j$ is
continuous and differentiable in $(0, 1)$, by the mean value theorem
there is a $c_j$, where $\min(x_j, y_j)< c_j < \max(x_j, y_j)$, such that
\beq{mvthe}
g_j'(c_j)=\frac{g_j(x_j)-g_j(y_j)}{x_j- y_j}
\eeq
for all $j$.
Substituting (43) into (42) and expressing (42)
in matrix form, we immediately prove (41).
\eproof
\end{proof}

The proof of Theorem 3 also needs the Poincar{\'{e}}-Miranda
Theorem, which
is a gereralization of the intermediate value theorem.
We quote the Poincar{\'{e}}-Miranda Theorem from [10].
Let $I^b=[0,1]^b$ be the $b$-dimensional cube of the Euclidean space $R^b$.
For each $i\le b$ denote
\[
I_i^- \defeq \{\bfx\in I^b : x_i=0\},\qquad I_i^+\defeq\{\bfx\in I^b : x_i=1\}
\]
the $i$-th opposite faces.
\bprop{PM} {\bf (Poincar{\'{e}}-Miranda Theorem)} Let
$\bff : I^b \to R^b$, $\bff=(f_1, f_2$, $\ldots$, $f_b)$, be a continuous map such that for
each $i\le b$, $f_i(I_i^-) \subset (-\infty, 0]$ and $f_i(I_i^+) \subset [0, +\infty)$.
Then, there exists a point $\bfc\in I^b$ such that $\bff(\bfc)=\bfzero$.
\eprop

Now we prove Theorem 3.
\begin{proof}[Theorem 3]
Now we analyze the first case in Theorem 3. We have shown that
fixed point $\bfone$ is attracting.  We now show that there is no other
fixed point in $[0, 1]^b$.  Suppose not.  Assume that there is another
distinct fixed point.  Denote it by $\bfx$.  From Lemma 1, we have
\beq{mateq1}
\bfone-\bfx=\bfJ(\bfc)(\bfone-\bfx).
\eeq
Since $g_i$ is a power series with non-negative coefficients, $g_i$
is monotonically increasing, differentiable and $g_i'$ is also increasing.
Thus,
\bear{upbound}
\bfJ(\bfc) &=& \phi(bq\bfH+(1-q)\bfone_{b\times b})\bfD\{g_1'(c_1),
g_2'(c_2), \ldots, g_b'(c_b)\} \nonumber \\
&\le& \phi(bq\bfH+(1-q)\bfone_{b\times b})\bfD\{g_1'(1),
g_2'(1), \ldots, g_b'(1)\} \\
&=& \bfJ(\bfone)\nonumber
\eear
component-wise.  Inequality (45) is due to the fact that
$\bfH$, $\bfone_{b\times b}$ and the two diagonal matrices are all non-negative.
Substituting the inequality above into (44),
we have
\[
\bfone-\bfx \le \bfJ(\bfone)(\bfone-\bfx).
\]
Substituting the last inequality repeatedly into itself, we have
\[
\bfone-\bfx\le \bfJ(\bfone)^n (\bfone-\bfx)
\to \bfzero,
\]
as $n\to\infty$, since the dominant eigenvalue of $\bfJ(\bfone)$
is strictly less than one.  We thus reach a contradiction to the assumption
that $\bfx$ is distinct from $\bfone$.

Now we consider the second case.  We first show that there exists
a point $\bfx$ in $[0, 1]^b$ such that
\[
\bfx-\bff(\bfx)\ge \bfzero.
\]
Denote such a point by $\bfeta$.  We choose
\beq{defeta}
\bfeta=\bfone-\epsilon\bfv,
\eeq
where $\epsilon$ is a small positive number and $\bfv$ is the eigenvalue
of $\bfJ(\bfone)$ associated with the dominant eigenvalue $\phi\lambda_1$.
For small $\epsilon$, we have
\[
\bff(\bfeta)=\bff(\bfone-\epsilon\bfv)\approx \bff(\bfone)-\bfJ(\bfone)(\epsilon\bfv)
=\bfone-\bfJ(\bfone)(\epsilon\bfv).
\]
It follows from the above equation that
\beq{eigvec}
\bfeta-\bff(\bfeta)\approx (\bfJ(\bfone)-\bfI)(\epsilon\bfv),
\eeq
where $\bfI$ is the $b\times b$ identity matrix.  Since $\bfv$ is an eigenvector
of $\bfJ(\bfone)$ associated with $\phi\lambda_1$, (47) reduces to
\[
\bfeta-\bff(\bfeta)=(\phi\lambda_1-1)\epsilon\bfv.
\]
Since $\phi\lambda_1 > 1$ and $\bfv > \bfzero$ entry-wise, we have
\beq{etaineq}
\bfeta-\bff(\bfeta)>\bfzero
\eeq
for some $\epsilon > 0$.

Next we shall show that (32) has another fixed point in
$[0, 1)^b$.
To apply Proposition 2, we transform system (32) by
changing variables.  That is, for any
$x_i\in [0, \eta_i]$, where $\eta_i$ is the $i$-th entry of $\bfeta$ defined
in (46).  We define
$y_i=x_i/\eta_i$, for $i= 1, 2, \ldots, b$.  Then, we define function $\bfF:
[0, 1]^b \to [0, 1]^b$, where the $i$-th entry of $\bfF$ is
\[
F_i(\bfy)=\eta_i y_i - f_i(\eta_1 y_1, \eta_2 y_2, \ldots, \eta_b y_b).
\]
We now show that for any $\bfy\in I_i^-$,
\begin{eqnarray*}
&&F_i(\bfy)\\
&&=-f_i(\eta_1 y_1, \eta_2 y_2, \ldots, \eta_b y_b) \\
&&=-(1-\phi)-\phi\left((bq+1-q)g_{h(i)}(\eta_{h(i)}y_{h(i)})+(1-q)\sum_{j\ne h(i)}
g_j(\eta_j y_j)\right) \\
&&\le 0,
\end{eqnarray*}
since $g_j(\eta_j y_j)\le 1/b$ for all $j$.  Next, consider $\bfy$ in $I_i^+$.
In this case,
\begin{eqnarray}
F_i(\bfy)&=&\eta_i-f_i(\eta_1 y_1, \ldots,\eta_{i-1} y_{i-1}, \eta_i,
\eta_{i+1} y_{i+1},\ldots, \eta_b y_b)\nonumber \\
&\ge&\eta_i -f_i(\eta_1,\ldots, \eta_{i-1},\eta_i, \eta_{i+1},\ldots, \eta_b)\label{Iplus1} \\
&\ge& 0,\label{Iplus2}
\end{eqnarray}
where (49) follows from the monotonicity of $g_j$ for all $j$, and
(50) follows from (48). From Proposition 2,,
$\bfF(\bfy)=\bfzero$ has a root in $[0, 1]^b$.  Equivalently, (32)
has a root in $[0, 1)^b$.  We denote this root by $\bfz$.

We now show that fixed point $\bfz$ is attracting.
From (41) since both $\bfone$ and $\bfz$ are fixed points, we have
\beq{mateq2}
\bfone-\bfz = \bfJ(\bfc)(\bfone-\bfz),
\eeq
where $z_i < c_i < 1$.  From (51), the unity is an eigenvalue
of $\bfJ(\bfc)$ and $\bfone-\bfz$ is the associated eigenvector.
Since $\bfJ(\bfc)$ is a positive matrix and $\bfone-\bfz$ is
a positive vector component-wise, by the Perron-Frobenius theorem,
the unity is the dominant eigenvalue of $\bfJ(\bfc)$ [14].
By the definition in (30), $g_i'$ is strictly increasing for all $i$.
It follows that $g_i'(c_i) > g_i'(z_i)$
and from (35) we have
\bear{Jlb}
\bfJ(\bfc) &=& \phi(bq\bfH+(1-q)\bfone_{b\times b})\bfD\{g_1'(c_1), g_2'(c_2),
\ldots, g_b'(c_b)\} \nonumber\\
&> & \phi(bq\bfH+(1-q)\bfone_{b\times b})\bfD\{g_1'(z_1), g_2'(z_2),
\ldots, g_b'(z_b)\} \nonumber\\
&=& \bfJ(\bfz)
\eear
component-wise.  Recall that we assume $q<1$.  With $\phi>0$, it is clear
that both $\bfJ(\bfc)$ and $\bfJ(\bfz)$ are irreducible matrices.
From Theorem 9 of [24](see also
[9]), (52) implies that
the spectral radius of $\bfJ(\bfz)$ is strictly less than
that of $\bfJ(\bfc)$.  This implies that $\bfz$ is an attracting
fixed point.

\eproof
\end{proof}

\bibliography{bibdatabase}
\bibliographystyle{apt}

\end{document}